\documentclass{article}
\usepackage{spconf,amsmath,amsfonts,graphicx}
\usepackage{tabularx}
\usepackage{cite}
\usepackage{svg}
\usepackage{balance}

\title{Improving Neural Diarization through Speaker Attribute Attractors and Local Dependency Modeling}
\name{David Palzer$^*$, Matthew Maciejewski$^\dag$, Eric Fosler-Lussier$^*$\thanks{This work was funded by The Johns Hopkins University Human Language Technology Center of Excellence.}}
\address{$^*$The Ohio State University, 
Computer Science and Engineering\\
$^\dag$The Johns Hopkins University, 
Human Language Technology Center of Excellence}
\begin{document}
%
\maketitle
\begin{abstract}

In recent years, end-to-end approaches have made notable progress in addressing the challenge of speaker diarization, which involves segmenting and identifying speakers in multi-talker recordings. One such approach, Encoder-Decoder Attractors (EDA), has been proposed to handle variable speaker counts as well as better guide the network during training. In this study, we extend the attractor paradigm by moving beyond direct speaker modeling and instead focus on representing more detailed `speaker attributes' through a multi-stage process of intermediate representations. Additionally, we enhance the architecture by replacing transformers with conformers, a convolution-augmented transformer, to model local dependencies.
Experiments demonstrate improved diarization performance on the CALLHOME dataset.

\end{abstract}
\vspace{-8pt}
\section{Introduction}
\label{sec:intro}




Speaker diarization is the activity of labeling an audio recording
with continuous segments by speaker identity. Diarization has a large range of applications, such as pre-processing audio for downstream tasks like source separation, information retrieval for pre-recorded audio, speaker-turn analysis for call recordings, and processing meetings \cite{tranter2006a}.
Diarization can also improve ASR performance for multi-speaker recordings (ICSI \cite{etin2006a}, AMI \cite{kanda2019a})
and acoustically noisy environments such as the family home (CHiME-5 \cite{kanda2018a}).

The lack of approaches that directly minimize diarization errors prompted the development of End-to-End Speaker Diarization (EEND) by Fujita et al. \cite{fujita2019a}. EEND is designed to address this challenge by optimizing diarization errors directly while using Permutation Invariant Training (PIT) to address speaker-label ambiguity \cite{yu2017a}. Notably, EEND, particularly the Self-Attentive EEND (SA-EEND) \cite{fujita2019b}, demonstrates the effectiveness of end-to-end training by surpassing traditional clustering-based methods. However, EEND has the limitation that the maximum number of speakers it can handle is predefined.

To tackle this limitation, Horiguchi et al. introduced Encoder-Decoder Attractor (EDA) models \cite{horiguchi2022a} that flexibly determine how many attractors are needed during the decoding of a speech embedding sequence.  This is achieved through compressing all speakers into a time-independent embedding and then separating out speaker attractors until the network predicts that all speakers have been produced. Every layer in the iterative network is required to predict a set of attractors.

In this study, we introduce {\em speaker attribute attractors} as a further improvement to the EEND-EDA architecture that disconnects layer wise prediction of speaker attractors from the iterative refinement, which provides 
a more robust way to condition deeper layers of the model. 
Additionally we integrate a conformer into our foundation model to inject local temporal dependencies, an approach that has shown promise on related audio processing tasks, which displays improvements in identifying speakers and finding speech.
\vspace{-8pt}
\section{Related Work}
Traditional diarization solutions
\cite{meignier2010a,shum2013a}
utilize clustering on speaker embeddings 
such as i-vectors \cite{dehak2011a}, d-vectors \cite{wang2018a}, and x-vectors \cite{garcia-romero2017a}. 
These speaker time embeddings are clustered using traditional ML algorithms, such as GMMs \cite{meignier2010a}, or hierarchical clustering \cite{sell2014a}.
 
Neural diarization techniques like EEND \cite{fujita2019a} employ an end-to-end neural network architecture that takes audio features as its input and produces the collective speech activities of multiple speakers as its output. The optimization of this network utilizes the entire recording, encompassing non-speech segments and instances of speaker overlap, with a primary focus on minimizing diarization errors.


Within the domain of speech separation, various techniques have been developed to handle mixtures involving a variable number of speakers. One set of methods follows the one-vs-rest approach, which is applied iteratively to separate speakers
\cite{kinoshita2018a,neumann2019a,takahashi2019a}. However, a significant drawback of this approach lies in the fact that calculations continue until all speakers are isolated, resulting in a linear increase in computational time as the number of speakers rises.

Another set of methods adopts attractor-based strategies, 
one example being the Deep Attractor Network (DANet) \cite{chen2017a}. 
While these approaches do not impose limits on the number of speakers during inference, they do require prior knowledge of the number of speakers. Anchored DANet \cite{luo2018a} effectively resolves these challenges; however, it still necessitates calculating dot products for all potential combinations of anchors and extracted embeddings during the inference phase. Consequently, its scalability with respect to the number of speakers is limited.


Further work on introducing label-dependency has been done by \cite{fujita2023neural} with the use of intermediate attractors. These intermediate attractors are added to each time frame as a weighted sum using intermediate diarization predictions before being passed to further layers. They find that this helps lower the missed speech rate, but also results in a higher false alarm rate. 
\vspace{-8pt}
\section{Methods}
This section describes the original EEND-EDA \cite{fujita2020a} and  EEND-EDA-deep \cite{fujita2023neural} architectures followed by our proposed improvements to both models.
\vspace{-8pt}
\subsection{EEND-EDA}
EEND-EDA modifies the original EEND network to use a stack of  $L$ transformers instead of LSTMs and to allow for an arbitrary number of speakers, which is not known a priori. 

\vspace{-8pt}
\begin{equation}
    E_l = \mathrm{Encoder}(E_{l-1}) \hspace{0.5in} (1 \leq l \leq L),
\end{equation}
where $E_l \in \mathbb{R}^{T\times D}$ is the sequence of embeddings of length $T$ , and dimension $D$.


The original EEND produced a static number of speakers, whereas with the inclusion of the EDA modification, EEND-EDA produces an attractor for each of $S$ speakers through the use of an LSTM and auxiliary existence loss for each produced attractor, $a_s$ where $(1\leq s \leq S)$.

\vspace{-8pt}
\begin{equation}
    A = \mathrm{EDA}(E_L), \mathrm{where}\  A = [a_1,...,a_S] \in \mathbb{R}^{S\times D}
\end{equation}

These speaker attractors are subsequently used to predict each speaker's frame-wise label by calculating the dot-product between each frame and each attractor. These speaker predictions are made concurrently and all frames are processed in parallel.

\vspace{-8pt}
\begin{equation}
    Y = \sigma(AE_L^T)
\end{equation}

Two of the downsides to this approach are the drop in performance as we produce more attractors, and the lack of label dependency between labels on our output.
\vspace{-8pt}
\subsection{Intermediate Attractors}
EEND-EDA with Intermediate Attractors was introduced by \cite{fujita2023neural}. They try to solve the lack of label dependency issue through creating intermediate speaker attractors and conditioning further layers on these attractors. This relaxes the conditional dependence between frames.

\vspace{-8pt}
\begin{equation}
    E_l = \mathrm{Encoder}(\hat{E}_{l-1}) \hspace{0.5in} (1 \leq l \leq L)
\end{equation}

They produce these intermediate attractors by injecting a shared EDA module between each transformer encoder which predicts the speaker activity at each layer. These shared intermediate EDA layers have an auxilary loss attached to them using the true speaker labels to help guide learning. These predicted speaker activities are then used as the attention weights of each speaker attractor for each frame. A shared linear projection layer is then used to project the frame-wise attractor conditioning before summing it with the current frame-embeddings and passing it to the next transformer encoder.

\vspace{-8pt}
\begin{equation}
    A_l = \mathrm{EDA}_l(E_l)
\end{equation}
\vspace{-8pt}
\begin{equation}
    \hat{E}_{l} = E_l + \sigma(E_lA_l^T)A_l^TW_l
\end{equation}


\vspace{-8pt}
\subsection{Attribute Attractors}
The task of directly modeling speakers, and conditioning further layers on these predictions, ties its expressiveness to how well each layer can individually solve the diarization problem. We instead propose a solution that relaxes this requirement in the form of attribute attractors, a non-autoregressive intermediate representation of our speakers that is more robust than direct speaker modeling.
These attribute attractors can be seen as a fixed size, over-segmentation of the embedding space of speakers. These attribute attractors are used to condition subsequent layers.

\begin{figure}
    \centering
    \includegraphics[width=\columnwidth]{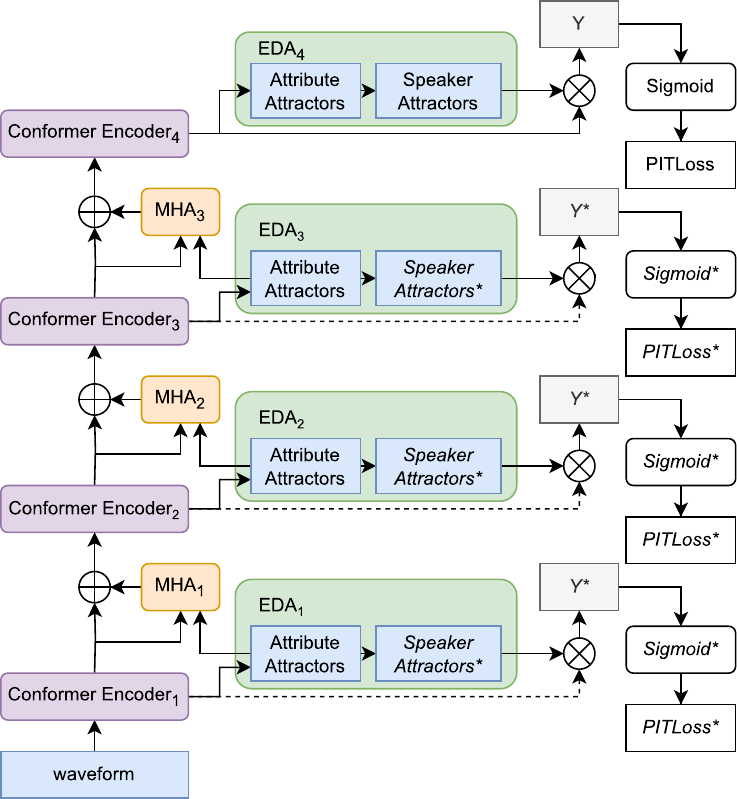}
    \caption{EEND with Attribute Attractors architecture. Layers and tensors in \textit{italics} with an asterisk (*) are removed after training and not used for inference.}
    \label{fig:model}
    \vspace{-4pt}
\end{figure}

\vspace{-8pt}
\begin{equation}
    A^a_l = \mathrm{EDA}^a_{l}(E_l)
\end{equation}
where $A^a \in \mathbb{R}^{N\times D} $ is our set of $N$ attribute attractors of dimension $D$. Note that $N$ is a hyper-parameter and is larger than $S$.

These attribute attractors are also used to produce auto-regressive speaker attractors in a similar fashion to the original EDA. These speaker attractors are used to predict speaker activity and require an auxilary loss to guide training. We found that simply using attribute attractors without predicting speaker activity did not work. Note: during inference the second half of the EDA module (speaker attractor production, and speaker activity prediction) is dropped for all but the last EDA layer as they are not used leading to all but the final EDA layer to be non-autogresssive, and a reduction in parameters and computation cost.

\vspace{-8pt}
\begin{equation}
    A^s_l = \mathrm{EDA}^s_l(A^a_l)
\end{equation}

Because we have now lost the ability to use intermediate predictions as attention weights for our attribute attractors, we modify the conditioning formulation to use multi-head-cross-attention instead of a simple scalar multiplication. This allows the network to learn which attributes it wants to condition each frame on. 

\vspace{-8pt}
\begin{equation}
    \hat{E}_l = E_l + \mathrm{MHA}(E_l,A^a_l)
\end{equation}
\vspace{-22pt}
\subsection{Conformer}
One downside of the existing attentive EEND models is their lack of temporal dependencies. In order to augment the network's ability to learn temporal dependencies we modify the architecture to use a conformer encoder instead of a transformer encoder. A conformer is a CNN augmented transformer and augments traditional self-attention with local dependencies. Conformers have been shown to outperform transformers on speech related tasks where predictions are highly dependent on local temporal information such as ASR \cite{gulati2020conformer}. While a relatively simple change to the model, we see an out-sized overall diarization improvement.
\vspace{-8pt}
\section{Experiments}
\vspace{-8pt}
\subsection{Data}
In following the example of the line of EEND studies, we prepare the CALLHOME, and Switchboard-2 (phase II, III) the same way, however we lack some of the data-sources that they have, namely Switchboard-2 Phase I and NIST Speaker Recognition Evaluation. Additionally we do not limit ourselves to 2-speaker mixtures as \cite{fujita2023neural} does due to our ability to handle arbitrary numbers of speakers. Our training set consists of simulated mixtures from the Switchboard-2 dataset and the CALLHOME training set, our validation set is the CALLHOME validation set, and our test set consists of the CALLHOME test set.

Features fed to each model consist of 23-dimensional log-scale Mel-filterbanks computed over 10 millisecond windows. We concatenate these mel frames using a window size of 15 and hop length of 10 to form 345-dimensional frames representing 100 milliseconds of audio. During training our data is augmented by selecting a 50s contiguous subset of the audio, and using the MUSAN \cite{musan2015} noise dataset for background sounds. In this work we do not use reverberation augmentation as found in \cite{landini2022simulated}. Additionally all models use the same batch size of 32, the AdamW optimizer, and are trained for 2000 epochs. We use the best performing models based on validation loss.

\begin{table*}[ht]
    \centering
    \begin{tabular}{lc|c|c|c|c|c|c|c|c}
    \hline
    \multicolumn{2}{|c|}{Model} & \multicolumn{2}{c|}{Parameters} & \multicolumn{4}{c|}{DER (\%)} & \multicolumn{2}{c|}{SAD (\%)}\\
    \hline
     & & Total & Free & DER & MS & FA & CF & MS & FA \\
        1. EEND-EDA \cite{fujita2020a} &&6.4M&6.4M& 9.96 & 5.40 & 1.36 & 2.81 & 3.85 & 0.87\\
        2. EEND-EDA-deep \cite{fujita2023neural} &&29.3M&17.0M& 8.50 & 4.43 & 1.31 & 2.76 & \textbf{3.15} & 0.85\\
        3. EEND with Attribute Attractors &&33.7M&33.7M& 7.87 & 4.18 & 1.42 & 2.27 & 3.28 & 0.84\\
        4. 3 + Conformer (Final system) &&35.3M&35.3M& \textbf{6.98} & \textbf{3.63} & 1.95 & \textbf{1.41} & 3.68 & \textbf{0.82}\\
        \hline\hline
        \multicolumn{9}{c}{Ablation study on changes required to get to EEND with Attribute Attractors}\\
        \hline
       5. 2 + Non-shared EDA &&29.3M&29.3M& 8.79 & 4.98 & 1.46 & 2.35 & 3.73 & 0.85\\
       6. 5 + Cross Attention && 32.1M & 32.1M & 8.48 & 4.52 & \textbf{1.30} & 2.66 & 3.23 & 0.86\\
       7. 6 + TransformerEDA &&32.9M&32.9M& 7.98 & 4.22 & 1.43 & 2.33 & 3.36 & 0.85\\
    \end{tabular}
    \caption{Diarization error rates (DER) for each model on the CALLHOME test set and component results}
    \label{tab:results}
    \vspace{-4pt}
\end{table*}

\begin{table}[]
\begin{tabular}{c|c|c|c|c|c|c}
\hline
\multicolumn{1}{|c|}{Layer} & \multicolumn{4}{c|}{DER (\%)}     & \multicolumn{2}{c|}{SAD (\%)} \\
\hline
      & DER   & MS    & FA   & CF   & MS         & FA         \\
1     & 26.07 & 10.86 & 8.40 & 6.81 & 8.34       & 0.81       \\
2     & 21.13 & 11.39 & 4.29 & 5.45 & 9.28       & 1.46       \\
3     & 15.81 & 6.90  & 3.34 & 5.57 & 6.15       & 0.93       \\
4     & 13.50 & 5.06  & 3.34 & 5.09 & 4.39       & 0.76       \\
5     & 12.26 & 5.09  & 2.51 & 4.67 & 4.21       & \textbf{0.60}       \\
6     & 9.99  & 4.32  & 2.59 & 3.08 & 4.11       & 0.73       \\
7     & 9.62  & 3.97  & 3.05 & 2.60 & 4.03       & 0.80       \\
8     & 8.07  & 4.56  & 1.86 & 1.65 & 4.50       & 0.80       \\
9     & 7.64  & 3.94  & 2.26 & 1.45 & 4.33       & 0.80        \\
10    & 7.13  & 3.67  & 2.04 & \textbf{1.41} & 3.89       & 0.78       \\
11    & 7.04  & 3.67  & \textbf{1.95} & 1.43 & 3.82       & 0.86       \\
\hline
Last  & \textbf{6.98}  & \textbf{3.63}  & \textbf{1.95} & \textbf{1.41} & \textbf{3.68}       & 0.82      
\end{tabular}
\caption{Diarization results for intermediate predictions of our final Conformer-based system across the CALLHOME test set.}
\label{tab:layers}
\end{table}
\vspace{-8pt}
\subsection{Model Configurations}
Baseline systems (1, 2) and final configurations (3, 4) correspond to the systems in Table~\ref{tab:layers}. Systems 5-7 indicate intermediate systems between systems 2 and 3 for ablation study purposes. Note: All models use 12 Encoder/EDA layers.\\
\textbf{EEND-EDA (1):} We re-implemented the baseline system \cite{fujita2020a}
to train on publicly available dataset splits.
This PyTorch implementation matches the original.\\
\textbf{EEND-EDA-deep with Intermediate Attractors (2):} This model reimplements
\cite{fujita2023neural} using intermediate attractors and intermediate auxiliary losses on all layers.\\
\textbf{Non-Shared EDA (5): } 
This first change to (2) unties
the EDA and projection layers. This untying does not impact processing speed, as the shared layers in the EEND-EDA-deep model are used $L-1$ times, whereas each of our $L-1$ intermediate EDAs are used once.\\
\textbf{+Cross-Attention Conditioning (6):} 
We replace weighted intermediate predictions with multi-head attention to 
produce weighted embeddings of unit length.
\\
\textbf{+TransformerEDA (7):} We
replace the 
LSTM based encoder in EDA  
with
an auto-regressive transformer
This also increases our training efficiency as we no longer need to randomly permute our time embeddings before processing them.\\
\textbf{+Attribute Attractors (3):} 
We further decouple
our intermediate predictions from the intermediate representation that they are guiding. Attribute attractors of size 256, produced by a non-autoregressive transformer, are used to predict intermediate speaker attractors with intermediate losses for training guidance. These attribute attractors are then used to condition further layers.\\
\textbf{+Conformer (4):} 
We swap our transformer backbone with conformers \cite{gulati2020conformer} to bias the transformers to pay attention to local context. 
These are convolution augmented transformers that have shown strong performance on speech tasks.\\
\vspace{-8pt}
\subsection{Metrics}
We calculate diarization error rate (DER) for each model as well as the DER components of missed speech (MS), false alarm (FA), and speaker confusion (CF). Additionally we compute the speech activity detection (SAD) components of missed speech, and false alarm.

\vspace{-8pt}
\balance
\subsection{Results}
Table \ref{tab:results} shows our DER and SAD results on the CALLHOME test set. 
Row 1 corresponds to the original EEND-EDA architecture and is our baseline for this study. It achieves a relatively low false alarm rate, but has an offsetting high missed speech rate while also struggling to disambiguate speakers. EEND-EDA-deep (Row 2) uses the intermediate attractor, self-conditioned architecture from \cite{fujita2023neural} and sees a large improvement on missed speech, with modest gains in false alarm and confusion rate. This model has the lowest SAD missed speech rate while not achieving the lowest DER missed speech rate.
This is due to the fact that SAD collapses all speech to speaking/non-speaking. Overlapped speech that is predicted as one speaker will count as a miss for DER but a hit for SAD.
Inclusion of attribute attractors (Row 3) improves our models ability to find speech and disambiguate speakers, while achieving a comparable false alarm rate.

Our conformer based model (Row 4) sees the lowest diarization error rate through a combination of large gains in missed speech rate and a trade off false alarm rate and confusion rate. The temporal aspect of the conformer helps discover speech, however this may be a potential downside of the features chosen and will be explored in the future. This trade-off is an overall improvement as our DER component rates are closer to parity. Additionally, Table \ref{tab:layers} shows that the DER for each layer throughout the model improves with the last layer performing best, which is in contrast with the EEND-EDA-deep original study \cite{fujita2023neural}.

The bottom half of Table~\ref{tab:results} provides an ablation study of the changes required to get to our conformer model. We see that unsharing the EDA layers is a losing proposition:  we see losses across the board. However, when we swap out the attention mechanism used to fold our attractors into our time embeddings, we see that we gain this performance back. We believe that the use of standard multi-head attention that is not tied to the current intermediate prediction, which allows an additional degree of freedom, is key.

The switch from an LSTM based EDA to an auto-regressive transformer delivers a modest gain in missed speech rate, however there is a trade-off in the false alarm rate being higher, but the rate of speaker confusion showing improvement.  This trade off appears to be from the fact that when a single person is speaking, our model is liable to predict two speakers instead of misattributing the speaker, this leads to a higher false alarm rate and a lower confusion rate.

\vspace{-8pt}
\section{Conclusion}
We created an end-to-end neural diarization architecture that uses attribute attractors to condition deeper layer of the network on our speakers. Additionally we updated parts of the model with modern attentive layers as well as introduced temporal inductive bias. We found that these changes resulted in improved diarization performance, particularly on the missed speech rate. Overall, these changes are beneficial to the EEND-EDA model.

\vfill\pagebreak



\bibliographystyle{IEEEbib}
\bibliography{main}

\begin{thebibliography}{10}

\bibitem{tranter2006a}
S.E. Tranter and D.A. Reynolds,
\newblock ``An overview of automatic speaker diarization systems,''
\newblock {\em IEEE Trans. on ASLP}, vol. 14, no. 5, pp. 1557–1565,, 2006.

\bibitem{etin2006a}
Ö. Çetin and E.~Shriberg,
\newblock ``Overlap in meetings: Asr effects and analysis by dialog factors,
  speakers, and collection site,''
\newblock in {\em Proc. MLMI}, 2006, p. 212–224.

\bibitem{kanda2019a}
N.~Kanda, C.~Boeddeker, J.~Heitkaemper, Y.~Fujita, S.~Horiguchi, K.~Nagamatsu,
  and R.~Haeb-Umbach,
\newblock ``Guided source separation meets a strong asr backend:
  Hitachi/paderborn university joint investigation for dinner party scenario,''
\newblock {\em INTERSPEECH}, p. 1248–1252, 2019.

\bibitem{kanda2018a}
N.~Kanda, R.~Ikeshita, S.~Horiguchi, Y.~Fujita, K.~Nagamatsu, X.~Wang,
  V.~Manohar, N.E.Yalta Soplin, M.~Maciejewski, S.J. Chen, A.S. Subramanian,
  R.~Li, Z.~Wang, J.~Naradowsky, L.P. Garcia-Perera, and G.~Sell,
\newblock ``Hitachi/jhu chime-5 system: Advances in speech recognition for
  everyday home environments using multiple microphone arrays,''
\newblock in {\em Proc. CHiME-5}, 2018, p. 6–10.

\bibitem{fujita2019a}
Y.~Fujita, N.~Kanda, S.~Horiguchi, K.~Nagamatsu, and S.~Watanabe,
\newblock ``End-to-end neural speaker diarization with permutation-free
  objectives,''
\newblock {\em INTERSPEECH}, p. 4300–4304, 2019.

\bibitem{yu2017a}
D.~Yu, M.~Kolbæk, Z.-H. Tan, and J.~Jensen,
\newblock ``Permutation invariant training of deep models for speaker
  independent multi-talker speech separation,''
\newblock in {\em ICASSP}, p. 241–245. 2017.

\bibitem{fujita2019b}
Y.~Fujita, N.~Kanda, S.~Horiguchi, Y.~Xue, K.~Nagamatsu, and S.~Watanabe,
\newblock ``End-to-end neural speaker diarization with self-attention,''
\newblock {\em ASRU}, p. 296–303, 2019.

\bibitem{horiguchi2022a}
Shota Horiguchi, Yusuke Fujita, Shinji Watanabe, Yawen Xue, and Paola Garcia,
\newblock ``Encoder-decoder based attractors for end-to-end neural
  diarization,''
\newblock {\em {IEEE}/{ACM} Transactions on Audio, Speech, and Language
  Processing}, vol. 30, pp. 1493--1507, 2022.

\bibitem{meignier2010a}
S.~Meignier,
\newblock ``Lium spkdiarization: An open source toolkit for diarization,''
\newblock in {\em CMU SPUD Workshop}. 2010.

\bibitem{shum2013a}
S.H. Shum, N.~Dehak, R.~Dehak, and J.R. Glass,
\newblock ``Unsupervised methods for speaker diarization: An integrated and
  iterative approach,''
\newblock {\em IEEE TASLP}, vol. 21, no. 10, pp. 2015–2028,, 2013.

\bibitem{dehak2011a}
N.~Dehak, P.J. Kenny, R.~Dehak, P.~Dumouchel, and P.~Ouellet,
\newblock ``Front-end factor analysis for speaker verification,''
\newblock {\em IEEE Trans. on ASLP}, vol. 19, no. 4, pp. 788–798,, 2011.

\bibitem{wang2018a}
Q.~Wang, C.~Downey, L.~Wan, P.Andrew Mansfield, and I.Lopez Moreno,
\newblock ``Speaker diarization with {LSTM},''
\newblock in {\em ICASSP}, p. 5239–5243. 2018.

\bibitem{garcia-romero2017a}
D.~Garcia-Romero, D.~Snyder, G.~Sell, D.~Povey, and A.~McCree,
\newblock ``Speaker diarization using deep neural network embeddings,''
\newblock in {\em Proc. ICASSP}, 2017, p. 4930–4934.

\bibitem{sell2014a}
G.~Sell and D.~Garcia-Romero,
\newblock ``Speaker diarization with plda i-vector scoring and unsupervised
  calibration,''
\newblock {\em SLT}, p. 413–417, 2014.

\bibitem{kinoshita2018a}
K.~Kinoshita, L.~Drude, M.~Delcroix, and T.~Nakatani,
\newblock ``Listening to each speaker one by one with recurrent selective
  hearing networks,''
\newblock in {\em ICASSP}, p. 5064–5068. 2018.

\bibitem{neumann2019a}
T.~Neumann, K.~Kinoshita, M.~Delcroix, S.~Araki, T.~Nakatani, and
  R.~Haeb-Umback,
\newblock ``All-neural online source separation, counting, and diarization for
  meeting analysis,''
\newblock in {\em ICASSP}, p. 91–95. 2019.

\bibitem{takahashi2019a}
N.~Takahashi, S.~Parthasaarathy, N.~Goswami, and Y.~Mitsufuji,
\newblock ``Recursive speech separation for unknown number of speakers,''
\newblock {\em INTERSPEECH}, p. 1348–1352, 2019.

\bibitem{chen2017a}
Z.~Chen, Y.~Luo, and N.~Mesgarani,
\newblock ``Deep attractor network for single-microphone speaker separation,''
\newblock in {\em ICASSP}, p. 246–250. 2017.

\bibitem{luo2018a}
Y.~Luo, Z.~Chen, and N.~Mesgarani,
\newblock ``Speaker-independent speech separation with deep attractor
  network,''
\newblock {\em IEEE/ACM TASLP}, vol. 26, no. 4, pp. 787–796,, 2018.

\bibitem{fujita2023neural}
Yusuke Fujita, Tatsuya Komatsu, Robin Scheibler, Yusuke Kida, and Tetsuji
  Ogawa,
\newblock ``Neural diarization with non-autoregressive intermediate
  attractors,''
\newblock in {\em ICASSP 2023}, 2023, pp. 1--5.

\bibitem{fujita2020a}
Yusuke Fujita, Shinji Watanabe, Shota Horiguchi, Yawen Xue, and Kenji
  Nagamatsu,
\newblock ``End-to-end neural diarization: Reformulating speaker diarization as
  simple multi-label classification,''
\newblock {\em ArXiv}, vol. abs/2003.02966, 2020.

\bibitem{gulati2020conformer}
Anmol Gulati, James Qin, Chung-Cheng Chiu, Niki Parmar, Yu~Zhang, Jiahui Yu,
  Wei Han, Shibo Wang, Zhengdong Zhang, Yonghui Wu, and Ruoming Pang,
\newblock ``Conformer: Convolution-augmented transformer for speech
  recognition,'' 2020.

\bibitem{musan2015}
David Snyder, Guoguo Chen, and Daniel Povey,
\newblock ``{MUSAN}: {A} {M}usic, {S}peech, and {N}oise {C}orpus,'' 2015,
\newblock arXiv:1510.08484v1.

\bibitem{landini2022simulated}
Federico Landini, Alicia Lozano-Diez, Mireia Diez, and Lukáš Burget,
\newblock ``From simulated mixtures to simulated conversations as training data
  for end-to-end neural diarization,'' 2022.

\end{thebibliography}

\end{document}